\documentclass[prd,superscriptaddress,unsortedaddress,twocolumn,showpacs,floatfix]{revtex4}

\usepackage[dvips]{graphicx}
\usepackage{dcolumn}
\usepackage{amsmath}
\usepackage{color}
\newcommand{\calo}{{\cal O}}
\begin{document}

\title{Dispersive analysis of $K_{L \mu 3}$  and $K_{L e 3}$
 scalar and vector form factors using KTeV data}

\newcommand{\UAz}{University of Arizona, Tucson, Arizona 85721}
\newcommand{\UCLA}{University of California at Los Angeles, Los Angeles,
                    California 90095}
\newcommand{\Campinas}{Universidade Estadual de Campinas, Campinas,
                       Brazil 13083-970}
\newcommand{\EFI}{The Enrico Fermi Institute, The University of Chicago,
                  Chicago, Illinois 60637}
\newcommand{\UB}{University of Colorado, Boulder, Colorado 80309}
\newcommand{\ELM}{Elmhurst College, Elmhurst, Illinois 60126}
\newcommand{\FNAL}{Fermi National Accelerator Laboratory,
                   Batavia, Illinois 60510}
\newcommand{\Osaka}{Osaka University, Toyonaka, Osaka 560-0043 Japan}
\newcommand{\Rice}{Rice University, Houston, Texas 77005}
\newcommand{\SaoPaolo}{Universidade de S\~ao Paulo, S\~ao Paulo, Brazil 05315-970}
\newcommand{\UVa}{The Department of Physics and Institute of Nuclear and
                  Particle Physics, University of Virginia,
                  Charlottesville, Virginia 22901}
\newcommand{\UW}{University of Wisconsin, Madison, Wisconsin 53706}

\affiliation{\UAz}
\affiliation{\UCLA}
\affiliation{\Campinas}
\affiliation{\EFI}
\affiliation{\UB}
\affiliation{\ELM}
\affiliation{\FNAL}
\affiliation{\Osaka}
\affiliation{\Rice}
\affiliation{\SaoPaolo}
\affiliation{\UVa}
\affiliation{\UW}

\author{E.~Abouzaid}      \affiliation{\EFI}
%%\author{T.~Alexopoulos}   \affiliation{\UW} %%Cascade polarazation, mag regen only
\author{M.~Arenton}       \affiliation{\UVa}
\author{A.R.~Barker}      \altaffiliation[Deceased.]{ } \affiliation{\UB}
\author{L.~Bellantoni}    \affiliation{\FNAL}
%%\author{A.~Bellavance}    \affiliation{\Rice} %%Remove Jan 07.  Keep on LFV paper.
\author{E.~Blucher}       \affiliation{\EFI}
\author{G.J.~Bock}        \affiliation{\FNAL}
\author{E.~Cheu}          \affiliation{\UAz}
\author{R.~Coleman}       \affiliation{\FNAL}
\author{M.D.~Corcoran}    \affiliation{\Rice}
%% \author{G.~Corti}         \affiliation{\UVa} %% only for pipiee papers
\author{B.~Cox}           \affiliation{\UVa}
\author{A.R.~Erwin}       \affiliation{\UW}
\author{C.O.~Escobar}     \affiliation{\Campinas}  %%Consider after Ke4 paper
\author{A.~Glazov}        \affiliation{\EFI}
%%\author{A.~Golossanov}    \affiliation{\UVa} %%Remove March 08
\author{R.A.~Gomes}       \affiliation{\Campinas}
\author{P. Gouffon}       \affiliation{\SaoPaolo}
%%\author{K.~Hanagaki}      \affiliation{\Osaka} %% Remove June 06
\author{Y.B.~Hsiung}      \affiliation{\FNAL}
%%\author{H.~Huang}         \affiliation{\UB}    %%Remove June 06
\author{D.A.~Jensen}      \affiliation{\FNAL}
\author{R.~Kessler}       \affiliation{\EFI}
\author{K.~Kotera}        \affiliation{\Osaka}
\author{A.~Ledovskoy}     \affiliation{\UVa}
\author{P.L.~McBride}     \affiliation{\FNAL}

\author{E.~Monnier}
   \altaffiliation[Permanent address ]{C.P.P. Marseille/C.N.R.S., France}
   \affiliation{\EFI}  %% Doug will ping him 

%%\author{K.S.~Nelson}     \affiliation{\UVa}  %Remove June 07
\author{H.~Nguyen}       \affiliation{\FNAL}
\author{R.~Niclasen}     \affiliation{\UB}
\author{D.G.~Phillips~II} \affiliation{\UVa}
%\author{H.~Ping}         \affiliation{\UW}  %%Remove Dec 07
%\author{V.~Prasad}       \affiliation{\EFI} %%case by case 832 only
%%\author{X.R.~Qi}         \affiliation{\FNAL} %%June 06
\author{E.J.~Ramberg}    \affiliation{\FNAL}
\author{R.E.~Ray}        \affiliation{\FNAL}
\author{M.~Ronquest}     \affiliation{\UVa}
\author{E.~Santos}       \affiliation{\SaoPaolo}
%%\author{J.~Shields}      \affiliation{\UVa} %% Jan 07.  Keep on pipig paper.
\author{W.~Slater}       \affiliation{\UCLA}
\author{D.~Smith}        \affiliation{\UVa}
\author{N.~Solomey}      \affiliation{\EFI}
\author{E.C.~Swallow}    \affiliation{\EFI}\affiliation{\ELM}
\author{P.A.~Toale}      \affiliation{\UB}
\author{R.~Tschirhart}   \affiliation{\FNAL}
%%\author{C.~Velissaris}   \affiliation{\UW}  %%Remove June 07?
\author{Y.W.~Wah}        \affiliation{\EFI}
\author{J.~Wang}         \affiliation{\UAz}
\author{H.B.~White}      \affiliation{\FNAL}
\author{J.~Whitmore}     \affiliation{\FNAL}
\author{M.~J.~Wilking}      \affiliation{\UB}
%\author{B.~Winstein}     \affiliation{\EFI}
\author{R.~Winston}      \affiliation{\EFI}
\author{E.T.~Worcester}  \affiliation{\EFI}
%\author{M.~Worcester}    \affiliation{\EFI} %%Removed Dec 07 by request of Ed
\author{T.~Yamanaka}     \affiliation{\Osaka}
\author{E.~D.~Zimmerman} \affiliation{\UB}
\author{R.F.~Zukanovich} \affiliation{\SaoPaolo}

\collaboration{KTeV Collaboration}\noaffiliation
\author{V\'eronique Bernard} \altaffiliation[Email:~bernard@ipno.in2p3.fr ]{ } 
\affiliation{Groupe de Physique Th\'{e}orique, IPN,
           Universit\'{e} de Paris Sud-XI, F-91406 Orsay, France}
\author{Micaela Oertel} \altaffiliation[Email:~micaela.oertel@obspm.fr] {} 
\affiliation{LUTH, Observatoire de Paris, CNRS, Universit\'e
  Paris Diderot, 5 place Jules Janssen, 92195 Meudon, France} 
\author{Emilie Passemar} \altaffiliation[Email:~passemar@itp.unibe.ch. Present address: 
IFIC, Universitat de Val\`encia - CSIC, Apt. Correus 22085, E-46071 Val\`encia, Spain] {} 
\affiliation{Albert Einstein Center for Fundamental Physics, Institute for theoretical physics,
University of Bern, Sidlerstr. 5, CH-3012 Bern, Switzerland} 
\author{Jan Stern} 
{}\altaffiliation[Jan Stern sadly passed away before this paper was completed.] {}  
\affiliation{Groupe de Physique Th\'{e}orique, IPN,
           Universit\'{e} de Paris Sud-XI, F-91406 Orsay, France}
\begin{abstract}
Using the published KTeV samples of $K_L\to \pi^{\pm} e^{\mp} \nu$
and $K_L\to \pi^{\pm} \mu^{\mp} \nu$ decays~\cite{KTeVe}, 
we perform a reanalysis of the scalar and vector form factors 
based on the dispersive parameterization~\cite{bops06,bops07a}.
We obtain phase space integrals 
$I^e_K     = 0.15446 \pm 0.00025$ and
$I^{\mu}_K = 0.10219 \pm 0.00025$.
For the scalar form factor parameterization,
the only free parameter is the normalized form factor value
at the Callan-Treiman point ($C$);
our best fit results in $\ln C=0.1915 \pm 0.0122$.  
We also study the sensitivity of $C$ to different 
parametrizations of the vector form factor.
The results for the phase space integrals and $C$ are
then used to make tests of the Standard Model.
Finally, we compare our results with lattice QCD calculations 
of $F_K/F_\pi$ and $f_+(0)$.
\end{abstract}


\maketitle

% \tableofcontents

% #######################
%
%       START 
%
% ######################

\section{Introduction}
\label{sec:intro}
Recently, much effort has been devoted to measure the vector
and scalar $K\pi$ form factors in semileptonic kaon decays 
in order to determine the phase space integrals.
These integrals, along with the kaon branching fractions,
%are needed 
allow to determine the CKM matrix element $|V_{us}|$. 
The scalar form factor ($f_0$) is difficult to measure because 
it is kinematically suppressed in $K_{e3}$ decays, 
and is therefore only measurable in $K_{\mu 3}$ decays,
which have contributions from both the scalar and vector form factors.
In addition, with the present experimental precision, 
only one $f_0$ model parameter can be accurately measured
(see e.g. \cite{franz}).
Until recently, these form factors were determined with
Taylor and pole parametrizations \cite{KTeVe,NA48e,ISTRAe, KLOEe}:
\begin{eqnarray}
\bar{f}_{+,0}^{Tayl}(t) & = & 
    1 + \lambda'_{+,0} \frac{t}{M_\pi^2} 
      + \frac{1}{2}\lambda''_{+,0} \left(\frac{t}{M_\pi^2}\right)^2
       \nonumber \\
   & & + \frac{1}{6}\lambda'''_{+,0} \left(\frac{t}{M_\pi^2}\right)^3 + \ldots 
\label{Taylor}
\end{eqnarray}
\begin{eqnarray}
    \bar{f}_{+,0}^{Pole}(t) &=& \frac{M_{V,S}^2}{M_{V,S}^2-t}~,
\label{pole}
\end{eqnarray}
where $t = (p_K - p_\pi)^2$ is the expansion parameter as a function
of the kaon and pion four-momenta, and $\bar{f}_{+,0}(t) \equiv
f_{+,0}(t)/f_{+,0}(0)$ are the normalized vector and scalar form
factors. The parameters measured in a fit to the data are
$\lambda'_{+,0}$ and $\lambda''_{+}$, the slope of the
form factors and the curvature of the vector one, 
and $M_{V,S}$, the mass of the vector and scalar
resonances. While the second order Taylor expansion has been used to
measure the vector form factor with sufficient precision, the scalar
form factor can only be determined using the first-order Taylor
expansion or the pole model. However,
% according to Ref. \cite{bops06}, 
clearly one has at least to know the curvature
to have a proper description of  $f_0(t)$ in the physical region
of $K_{\ell 3}$-decays.
%
%Indeed it was pointed out that since the measured slope parameter
%using the linear parametrization can only be an upper bound for the
%true mathematical slope due to the positive curvature, ignoring the
%higher order terms induces a systematic error on the value of the
%slope parameter, which is non negligible, see Ref. \cite{bops06}. 
%
After results were reported based on the Taylor and pole 
parametrizations, 
a form factor parametrization based on conformal mapping 
was discussed in Ref. \cite{bechhill}
in the context of $B \to \pi l \nu$ 
to improve the convergence of the series and to
give rigorous bounds on its coefficients. 
%of the series. 
This parametrization was applied to the $K_{\ell 3}$ case in 
Ref. \cite{hill}, and recently used by the KTeV collaboration
to reanalyze their $K_{e3}$ data \cite{KTeVz}.

%Dispersion relations have also been used to describe the shape of the
%form factors
%cite{bops06,JOP,Bachir:08,Jamin:2008qg,Boito:2008fq}. 
%% the partial knowledge of the 
%measured low energy $K \pi$ phase shifts \textcolor{blue}{\cite{mouss,Estabrooks,Aston:1987ir}}.  
%As an alternative approach
%the dispersive parametrization in Ref. \cite{bops06,bops07a} 
%has been specially constructed
%for being used in the analysis of Kl3 measurements. It 
%has the advantage
%to account for the correlation between the slope and the curvature,
%by using low energy $K \pi$ phase shifts. 
%It involves  only one free parameter for both the scalar
%and vector form factor to be determined from the existing data sample.}
%Whereas in Refs. \cite{JOP,Bachir:08,Jamin:2008qg,Boito:2008fq}, the shape of the form factors 
%is completely determined from high energy data, the approach in Ref. \cite{bops06} 
%is somewhat different. 
%The form factors are expressed in terms 
%of only one free parameter 
%to be determined from the existing $K_{\ell 3}$ data sample. 
As an alternative approach, the dispersive parametrization in Refs. \cite{bops06,bops07a} 
has the advantage to account for the correlation between the slope and the curvature,
by using low energy $K \pi$ phase shifts \cite{mouss,Estabrooks,Aston:1987ir}. 
It involves only one free parameter for
both the scalar and vector form factors to be determined from the existing data sample.
%Indeed, this parametrization has the advantage
%to account for the correlation between the slope and the curvature,
%by using low energy $K \pi$ phase shifts \cite{mouss,Estabrooks,Aston:1987ir}.} 
The sole scalar form factor parameter is $C$, 
the value of the normalized scalar form factor at the 
Callan-Treiman (CT) point, 
$t \equiv \Delta_{K \pi}=m_K^2-m_\pi^2$, 
the difference of kaon and pion masses squared. 
Once $C$ is determined, the shape of the scalar form factor 
is known with a high precision in the physical region
and somewhat beyond.
%% It therefore reduces the theoretical uncertainties in the analysis.  
The choice of this particular parameter $C$ is guided by the 
existence of the Callan-Treiman theorem~\cite{Dashen:1969bh} 
which predicts its value in the
$\mathrm{SU}(2)\times \mathrm{SU}(2)$ chiral limit. 
For physical quark masses,
\begin{equation}
   C\equiv \bar f_0(\Delta_{K\pi})  =
    \frac{F_{K^+}}{F_{\pi^+}}\frac{1}{f_{+}^{K^0\pi^-}(0)} + \Delta_{CT}~.
%   \vspace{-0.1cm} 
     \label{C}
\end{equation}
where $F_{\pi}$ and $F_{K}$ are the pion and kaon decay constants, 
respectively, and $\Delta_{CT}$ is a correction of order 
$\calo \left( m_{u,d}/4 \pi  F_\pi \right)$ 
arising from non zero quark masses $m_u,m_d$.  
This correction has been evaluated within 
Chiral Perturbation Theory (ChPT) and is small enough
that the right-hand side of Eq.~(\ref{C}) can be determined
%from experimental inputs 
%assuming the SM EW 
%couplings of quarks}} 
with sufficient accuracy as discussed in \S~\ref{sec:discussion} 
to compare with $C$ measured 
in
%from the scalar form factor 
%{\textcolor{blue} {by a fit to the 
$K_{\mu 3}$-decays. %data}. 
Thus apart from the determination of $|V_{us}|$,
which is used to test the unitarity of the CKM matrix within the 
Standard Model (SM), 
a measurement of the scalar form factor at the Callan Treiman point 
provides another interesting test of the SM, namely 
a test of the couplings of light quarks to $W$. 
Another interest in the experimental determination of the shape 
of the $K \pi$ scalar form factor is the possibility of 
determining low energy constants which appear in ChPT~\cite{be08}.

The NA48 \cite{NA48mu} and KLOE \cite{KLOE} collaborations
have reanalyzed their data with the dispersive parameterization
\cite{bops06}. The values of $C$ obtained in these two experiments 
differ by $2.1\sigma$. Here we present a similar reanalysis of the 
KTeV data~\cite{KTeVe} leading to an improvement on the precision on the 
determination of the form factors compared with the 
previous KTeV results~\cite{KTeVe,KTeVz}.  
Since the vector and scalar form factors are correlated,
alternative parametrizations for the vector form factor
are studied to probe the robustness of the scalar form factor result.

The paper is organized as follows. 
In \S~\ref{sec:analysis} we present the results of the 
dispersive analysis of the KTeV data. 
In \S~\ref{sec:parametrisations} we discuss the
correlations between the vector and the scalar form factor. 
\S~\ref{sec:discussion} is devoted to a discussion of
different applications of our results, in particular the test
of the SM.  We summarize in \S~\ref{sec:summary}.

%%%%%%%%%%%%%%%%%%%%%%%%%%%%%%%%%%%%%%%%%%%%%%%%%%%%%%%%%%%%%%%
\section{Dispersive Analysis of KTeV Semileptonic Data}
%%%%%%%%%%%%%%%%%%%%%%%%%%%%%%%%%%%%%%%%%%%%%%%%%%%%%%%%%%%%%%%
\label{sec:analysis}
Assuming $\bar{f}_0(t)$ is never equal to zero, 
the dispersive representation 
for the normalized scalar form factor reads
\begin{eqnarray}
\label{Dispf}
\bar f_0(t)&=&\exp\Bigl{[}\frac{t}{\Delta_{K\pi}}(\mathrm{ln}C- G(t))\Bigr{]}~,  \\ 
G(t)&=&\frac{\Delta_{K\pi}(\Delta_{K\pi}-t)}{\pi}
\label{G}  \\
& & \times \int_{(m_K+m_\pi)^2}^{\infty}
\frac{ds}{s}
\frac{\phi_0(s)}
{(s-\Delta_{K\pi})(s-t-i\epsilon)}~. \nonumber
\end{eqnarray}
Note that $C$ is here the only free parameter. $\phi_0(s)$ represents
the phase of the form factor: following Watson's theorem
\cite{Watson}, this phase is equal to the 
%experimentally determined 
$K\pi$ scattering phase
within the elastic region. 
In writing Eq.~(\ref{Dispf}), two subtractions have been made 
to minimize the unknown high energy  
contribution to the 
dispersive integral, Eq.~(\ref{G}).
%from the inelastic part at high energy, entering $\phi_0$. } 
%,that come from the inelastic component 
%in the high-energy region.
The two subtraction points have been taken at $t = 0$ 
and at the CT point to take advantage of the CT theorem, 
Eq.~(\ref{C}).  The resulting function $G(t)$ in Eq.~(\ref{G})
does not exceed 20\% of the expected value of ln$C$;
since theoretical uncertainties on $G(t)$ are $\sim 10$\% its value,
the corresponding uncertainty on $\ln C$ is then a 
few percent of its value.

\begin{table*}[t]
\begin{center}
\begin{tabular}{|l|c|c|c|}
\hline
                 &  $K_{e 3}$ only  &  $K_{\mu 3}$  only & $K_{e 3}$ and $K_{\mu 3}$ Combined \\
\hline
$\Lambda_+\times 10^3$ & 25.17 $\pm$ 0.62 & 24.57 $\pm$ 1.10 & 25.09 $\pm$ 0.55 \\
$\ln C$                & - & 0.1947 $\pm$ 0.0140 & 0.1915 $\pm$ 0.0122 \\
$\rho (\Lambda_+, \ln C)$ & - & -0.557 & -0.269 \\
$\chi^2/{\rm dof}$       & 66.6/65 & 193/236& 0.48/2     \\
\hline 
\hline
$\lambda'_+\times 10^3$ & 25.17 $\pm$ 0.62 & 24.57 $\pm$ 1.10 & 25.09 $\pm$ 0.55 \\
$\lambda''_+\times 10^3$ & 1.22 $\pm$ 0.07 & 1.19 $\pm$ 0.07 & 1.21 $\pm$ 0.08 \\
$\lambda_0\times 10^3$ &- & 13.22 $\pm$ 1.39 & 12.95 $\pm$ 1.17\\
$\lambda'_0\times 10^3$ &- & 0.59 $\pm$ 0.05 & 0.58 $\pm$ 0.05 \\
\hline
\hline
 $I_{K}^{e}$   & 0.15450 $\pm$ 0.00028 &   0.15416 $\pm$ 0.00060 & 0.15446 $\pm$ 0.00025   \\ 
 $I_{K}^{\mu}$   & -     &   0.10207 $\pm$ 0.00032 &  0.10219 $\pm$ 0.00025 \\ \hline 
 $I_{K}^{\mu}$/$I_{K}^{e}$ & - & 0.6621 $\pm$ 0.0018 & 0.6616 $\pm$ 0.0015 \\ \hline 
%\hline
% $R_{\mu/e}$    & - & 0.6650 $\pm$ 0.0034 & 0.6645 $\pm$ 0.0033 \\
% \hline

\end{tabular}
\end{center}
\caption{\label{tab:disp}{\it Results of the analysis of the KTeV $K_{L e 3}$ and $K_{L \mu 3}$ data
using a dispersive parameterization for the vector and scalar form factors. 
$\Lambda_+$ and $\ln C$ are the parameters of the fit used to calculate the 
slopes, curvatures and the phase space integrals. The uncertainties correspond to the total ones, adding the statistic, 
the systematic as well as the theoretical ones in quadrature, see the text for
more details.}}
\end{table*}

The dispersive representation of the vector form factor is constructed
in a similar manner. Since there is no analog of the CT theorem in this
case, the two subtractions are performed at $t = 0$. 
The normalized vector form factor is
\begin{eqnarray}
\label{Dispfp}
\bar f_+(t) & = & 
     \exp\Bigl{[}\frac{t}{m_\pi^2}\left(\Lambda_+ + H(t)\right)\Bigr{]}~, \\
 H(t) & = & 
     \frac{m_\pi^2t}{\pi} \int_{(m_K+m_\pi)^2}^{\infty}
     \frac{ds}{s^2}\frac{\phi_+ (s)}{(s-t-i\epsilon)}~,
\label{H}
\end{eqnarray}
where $\Lambda_+ \equiv m_\pi^2 d \bar f_+(t)/dt|_{t=0}$ and
$\phi_+(s)$ is the phase of the vector form factor. 
As in the case for the scalar form factor,
information on the $K\pi$ phase shifts in the elastic region 
is used to determine $\phi_+(s)$.  
The main contribution to $\phi_+(s)$
is the dominant $K^*(892)$ resonance.
The extrapolation of the $K\pi$ phase shift
data down to threshold is done here following a Gounaris-Sakurai
construction based on the $K^*(892)$ and exhibiting the correct
threshold behavior and the correct properties of 
analyticity and unitarity. 
The value of $H(t)$ represents at most 20\% of the value of 
$\Lambda_+$ such that the latter can be measured with 
high precision.
For more details on the dispersive representations,
see Refs. \cite{bops06} and \cite{bops07a}. 

In Ref. \cite{bops07a}, a thorough discussion of the different sources of
theoretical uncertainties of the dispersive representations can be
found. They include the error on the low energy $K \pi$ phase shifts 
and an estimate of the uncertainties due to the unknown high energy
behaviour of the phases $\phi_0(s)$ and $\phi_+(s)$.
The corresponding error-bands, $\delta G(t)$ and $\delta H(t)$, 
are used in this analysis
to propagate uncertainties on $\Lambda_+$ and $\ln C$. 

The analysis of the KTeV data is done using their $K_{L e3}$ and $K_{L\mu 3}$ 
samples with $1.9 \times 10^6$ and $1.5 \times 10^6$ events,
respectively after selection requirements.  These samples were
collected in a special run in which the beam intensity
was lowered by a factor of ten compared to that used to measure
$\epsilon'/\epsilon$.  The laboratory-frame kaon energies are 40-160 GeV
(mean is 70 GeV), and the momenta of charged particles are measured with
much better than 1$\%$ precision. Muons are identified with a large
scintillator hodoscope behind 3 meters of steel. Electrons and pions
are identified primarily by ratio of energy deposited in the cesium
iodide calorimeter ($E$) to the momentum measured in a magnetic
spectrometer ($p$); $E/p \sim 1$ for electrons, and $E/p < 1$ for
pions. In addition to using the KTeV data, we also use the KTeV Monte
Carlo (MC) to correct for the detector acceptance that results in a
non-uniform sampling of the $K_{\ell 3}$ Dalitz plot. 

The results of the dispersive analysis are given in Table
\ref{tab:disp}. The associated slope and curvature are also given,
based on Taylor expansions of Eqs.~(\ref{Dispf}) and (\ref{Dispfp})
using the best-fit values of $\ln C$ and $\Lambda_+$, respectively.
A combined $K_{e 3}$ and $K_{\mu 3}$ dispersive analysis leads to
\begin{equation}\label{eq:lnc}
\begin{array}{lcl}
\Lambda_+ &=& 0.02509 \pm 0.00035_{\rm stat} \pm 0.00027_{\rm syst} \pm 0.00033_{\rm th}\\
          &=& 0.02509 \pm 0.00055,\\
\ln C      &=& 0.1915 \pm 0.0078_{\rm stat} \pm 0.0086_{\rm syst} \pm 0.0038_{\rm th} \\
          &=& 0.1915 \pm 0.0122.
\end{array}
\end{equation}
The phase space integrals are then given by:
\begin{equation} \label{eq:phase}
\begin{array}{lcl}
I_K^e &=&   0.15446 \pm 0.00019_{\rm stat} \pm 0.00015_{\rm syst} \pm 0.00008_{\rm th} \\
%       0.00019_{\rm th} \pm 0.00011_H\\
       &=&   0.15446 \pm 0.00025,\\
I_K^{\mu}&=& 0.10219 \pm 0.00017_{\rm stat}  \pm 0.00017_{\rm syst} \pm 0.00005_{\rm th} \\
%\pm 0.00014_{\rm th} \pm 0.00008_H \pm 0.00006_G \\ 
       &=& 0.10219 \pm 0.00025. \\
\end{array}
\end{equation}
Systematic uncertainties are estimated following the prescription
of Ref.~\cite{KTeVz}, by scaling the ratio of
systematic-to-statistical uncertainties for the pole
model in Table~1 of Ref.~\cite{KTeVe}
(see Eq. (15) of Ref. \cite{KTeVz}). 
Statistical, systematic and theoretical uncertainties are added 
in quadrature to give the total uncertainty. 
To estimate the theoretical error on $\ln C$ and $\Lambda_+$ 
induced by uncertainties on the functions $G(t)$ and $H(t)$ 
entering the dispersive representations, 
we perform fits using $G(t) \pm \delta G(t)$
and $H(t) \pm \delta H(t)$.
The function $G(t)$ is positively correlated with $\ln C$,
and $H(t)$ is negatively correlated with $\Lambda_+$;
these correlations lead to reduced uncertainties
in the phase-space integrals ($I_K$).

Table~\ref{tab:disp} provides also values of the phase space
integrals ratio, $I^{\mu}_K/I^e_K$. Note that for them
the estimated total uncertainty takes into account  correlation
due to the common vector form factor $f_+(t)$ which reduces
the uncertainty.

After subtracting the common theoretical uncertainties,
our result for  $\ln C$ is consistent with the KLOE result, 
$\ln C=0.2038(246)$ \cite{KLOE}, and it is $2.6\sigma$ 
larger than the NA48 result, $\ln C=0.1438(140)$ \cite{NA48mu}.
For the previous form factor fits from KTeV \cite{KTeVe},
the phase-space integrals
($I_K^e = 0.15350(105), I_K^\mu = 0.10165(80)$)
are in good agreement with the dispersive results above,
but the precision is limited by modeling uncertainties 
that are twice as large as the statistical errors.
The large modeling uncertainty in Ref. \cite{KTeVe} is based on the 
difference between using the pole model 
($I_K^e = 0.15445(23_{stat}), I_K^\mu = 0.10235(22_{stat})$)
and the quadratic model
($I_K^e = 0.15350(44_{stat}), I_K^\mu = 0.10165(39_{stat})$),
where the uncertainties are statistical only;
the pole model result agrees very well with the
result above based on the dispersive analysis.
For the $z$-parameterization \cite{KTeVz}
($I_K^e = 0.15392(48)$), the  theoretical uncertainty
is slightly larger than that from the dispersive analysis,
and therefore the $I_K^e$-discrepancy may be significant.
To investigate this difference, several  Monte Carlo samples
were generated using input form factors from the result 
of the dispersive fit and subsequently analysed using $z$-parameterization. 
Based on this study, the $I_K$ integrals obtained with the
$z$-parameterization reproduce on average the input value, and the
difference between the $z$ and dispersive parameterization observed
for the KTeV data is consistent with a $1.8\sigma_{\rm stat}$
fluctuation.

%%%%%%%%%%%%%%%%%%%%%%%%%%%%%%%%%%%%%%%%%%%%%%%%%%%%%%%%%%%%%%%%%%%%%%%%%%%
%The results for the phase space integrals are compatible within the
%uncertainties with the previous analysis of the KTeV data: $I_K^e =
%0.15350(44), I_K^\mu = 0.10165(39)$ (quadratic model \cite{KTeVe}),
%$I_K^e = 0.15445(23), I_K^\mu = 0.10235(22)$ (pole model \cite{KTeVe}), 
%$I_K^e = 0.15392(48)$ (z-paramete\-rization \cite{KTeVz}).  Note
%that the slope and the curvature of the vector form factor and
%consequently $I_K^e$ are in perfect agreement with the values computed
%from the pole fits from the $K_{e3}$ measurements of KTeV, KLOE and
%NA48. 
%Furthermore $\ln C$ is consistent with the KLOE result, 
%$\ln C=0.2038(246)$ \cite{KLOE} and larger than the NA48 result, 
%$\ln C=0.1438(140)$ \cite{NA48mu}.

%%%%%%%%%%%%%%%%%%%%%%%%%%%%%%%%%%%%%%%%%%%%%%%%%%%%%%%%%%%%%%%%%%
\section{Comparison of Different Form Factor Parameterizations}
\label{sec:parametrisations}
%%%%%%%%%%%%%%%%%%%%%%%%%%%%%%%%%%%%%%%%%%%%%%%%%%%%%%%%%%%%%%%%%%

As pointed out in the introduction, a main advantage of the dispersive
parameterization is the possibility of determining the value of the
scalar form factor at the Callan-Treiman point, thus allowing for a
test of the SM. At present, only the dispersive parameterization
makes it possible 
to determine the scalar form factor at this point, which lies far beyond the 
endpoint of the physical region, with reasonable precision.
Since there is a large correlation between $\ln C$ and the slope of the
vector form factor (see $\rho (\Lambda_+,\ln C)$ in Table \ref{tab:disp}),
however, it is important to study the sensitivity of $\ln C$ to the choice
of parameterization for the vector form factor.

To investigate this sensitivity, 
we have fit the KTeV data using the dispersive parameterization for the
scalar form factor (with $\ln C$ as a free parameter) and
four different parameterizations for the vector
form factor:
\begin{itemize}
\item the dispersive parameterization Eq.~(\ref{Dispfp})
\item the pole parameterization Eq.~(\ref{pole})
%[*** OMIT which  assumes that the vector 
%form factor is dominated by a resonance of mass $M_V$ which is the only
%parameter of the fit. 
%:
%\begin{equation}
%\bar{f}_+(t) = \frac{1}{1-t/m_V^2} \,\, .
%\end{equation}
%The result of the pole fit gives $M_V$ in good agreement with the $K^*$
%resonance, see Table 2 showing that the assumption of $K^*$ dominance in a 
%certain energy region is consistent with the KTeV data. ***OMIT]
%
\item the quadratic (second-order) Taylor expansion Eq.~(\ref{Taylor})
\item the z-parameterization
\begin{equation}
F_{+}(t) = F_{+}(t_0) \frac{\phi_{+}(t_0,t_0,Q^2)}{\phi_{+}(t,t_0,Q^2)}
\sum_{k=0}^\infty a_k (t_0,Q^2) z(t,t_0)^k~, \nonumber 
\end{equation}
\vspace{-0.5cm}
\begin{equation}
\bar{f}_{+}(t) = F_{+}(t)/F_{+}(0)~,
\label{fzpar}
\end{equation}
based on a conformal mapping of $t$ onto the
variable $z$ with 
\begin{equation}
z(t,t_0) \equiv \frac{\sqrt{t_+-t} -
  \sqrt{t_+-t_0}}{\sqrt{t_+-t}+\sqrt{t_+-t_0}}~.
\end{equation}
In Eq. (\ref{fzpar}) we have used the notation from Ref. \cite{hill}. 
%[*** OMIT 
%\textcolor{blue}{In this
%paper, it has been shown that for a specific choice of $\phi_+$, the
%sum of the expansion coefficients is bounded based on unitarity and
%the total rate of $\tau \to K \pi \nu$ decays, a significant advantage
%compared to the quadratic parameterization.  For example one
%of the choices in Ref. \cite{hill} leads to $\sum_{k=0}^\infty a_k^2 <13$.
%Comparing the experimental precision to this bound, the best total
%uncertainties for the $z$-parametrization are obtained for a two
%parameter fit~\cite{KTeVz}~\footnote{Once these two parameters are
%  determined one can calculate the slope and the curvature, see
%  table~\ref{tab:Ke3formfactors}, \ref{tab:Kl3formfactors}. All higher
%  order terms in the Taylor expansion are functions of these two
%  parameters and thus are directly related to the slope and the
%  curvature.}.} 
%***OMIT]

\end{itemize}

The results of these fits are summarized in 
Tables \ref{tab:Kl3formfactors} and \ref{tab:Ke3formfactors}. 
The ``dispersive'' results are taken from 
Section~\ref{sec:analysis}. All of the fits have good $\chi^2 /dof$. 
Interestingly, the pole and dispersive parameterizations result in
very similar values for $\ln C$, while the quadratic and z-parameterization 
results are similar. This can in fact be easily understood from the Appendix which presents
a detailed investigation of the correlations between parameters in the different
parameterizations. All of the $\ln C$ results are consistent within $2\sigma_{\rm stat}$,
as can be deduced from the difference of the $\chi^2$ (4 units per one
degree of freedom change) and by estimating the uncertainty of the
difference as a difference of the uncertainties in quadrature: $\ln
C|_{\rm disp/disp} - \ln C|_{z/{\rm disp}} = 0.026 \pm 0.013$.
% The Appendix presents a detailed investigation of the 
%correlations between parameters in the different parameterizations.

\begin{table*}
\begin{center}
    \begin{tabular}{|l||c|c|c|c|c|}
    \hline
& \multicolumn{4}{c|}{}  \\
\vspace{-0.1cm}
     & \multicolumn{4}{c|}{Parameterization Vector FF/Scalar FF}  \\
\vspace{-0.1cm}
& \multicolumn{4}{c|}{}  \\
  \hline  
        Results    & disp.(I)/disp.  &  pole(I)/disp. &  quad (II)/disp.  & z-param.(II)/disp.\\ \hline  \hline 
 Fit param.  $v_i$                         &$\Lambda_+ =24.57(83)$   & $M_V =890.00 $ &  $\lambda_+^{'}
   =17.5(3.4)$                  &   $a_1 = 1.057(63)$ \\
& & $(13.00) \mathrm{MeV}$ &  $\lambda_+^{''}
                                         = 4.3(1.4)$ & $a_2 = 3.9(3.2)$ \\
&&&&\\
 $\lambda'_+$ &24.57(83)& 24.59(72) & 17.5(3.4)        & 20.00(2.60)\\
 $\lambda^{''}_+ $ & 1.19(4)& 1.21(7)& 4.3(1.4)        & 2.5(6)\\ \hline \hline
Fit param. $\ln C$ &0.1947(91) & 0.1944(93) &  0.169(16)       & 0.170(16)  \\
&&&&\\
$\lambda_0 $& 13.22(78)& 13.20(79) & 11.03(1.37)      &  11.11(1.37)\\
$\lambda_0^{'} $  & 0.59(2)& 0.59(2) & 0.54(3)      &  0.54(3) \\
\hline
$\rho(v_i,\ln C)$    & -0.557 &  0.588  &  0.707
&  0.477    \\
& -&-&-0.819&-0.766\\
 \hline  
$\chi^2$/dof   & 193/236  & 193/236  & 189/235                  &  189/235 \\ \hline
%\end{tabular}
     \end{tabular}
   \end{center}
   \caption{\it Results of the fit to $K_{L\mu 3}$ data using different
     parameterizations for the vector form factor and the dispersive
     one for the scalar form factor. The uncertainties are only statistical.
$\Lambda_+$, $\lambda'_{+}$, $\lambda''_+$,  $\lambda_{0}$ and $\lambda'_0$  are in units $10^{3}$.}
    \label{tab:Kl3formfactors}
 \end{table*}

\begin{table*}[t]
\begin{center}
    \begin{tabular}{|l||c|c|c|c|c|}
    \hline
& \multicolumn{4}{c|}{}  \\
\vspace{-0.1cm}
     & \multicolumn{4}{c|}{Parameterization Vector FF}  \\
\vspace{-0.1cm}
& \multicolumn{4}{c|}{}  \\
  \hline   
    Results &  dispersive (I) &    pole (I) &   quadratic (II) & z-param. (II)
\\ \hline  \hline          
   Fit param.     & $\Lambda_+ =25.17(38)$   &  $M_V= 881.03$&      
$\lambda_+^{'}=
   21.67(1.59)$               &   $a_1 = 1.023(28)$    \\            
&&$(5.91) \mathrm{MeV}$&$\lambda_+^{''} = 2.87(66)$                       &
    $a_2 = 0.75 (1.58)$  \\
&&&&\\
 $\lambda'_+  $ & 25.17(38)&
 25.10(41) & 21.67(1.59)        & 22.69(1.20)  \\
 $\lambda''_+ $ & 1.22(4)&
 1.26(3) & 2.87(66)       & 1.93(30)  \\ \hline \hline
Correlation      &  -  &  -  &  -0.96            &
-0.064 \\ \hline
$\chi^2/$dof   &66.6/65 & 66.3/65 & 62.2/64      &  62.3/64\\ \hline
     \end{tabular}
\caption{{\it{Results of the fit to $K_{Le3}$ data using different 
parameterizations with only the statistical uncertainties. Pole, quadratic and $z$-parameterization results are
from Refs. \cite{KTeVe} and \cite{KTeVz}.
$\Lambda_+$, $\lambda'_+$ and $\lambda''_+$ are in units $10^{3}$. \label{tab:Ke3formfactors}}}}    \end{center}
\end{table*}

A similar level of agreement between the different parametrizations is
observed for the $I_K$ integrals.  For example for $K_{L\mu3}$ data, using the
$z$-parameterization with two parameters for each of the vector and 
scalar form factors, one
obtains $I_K^e|_{z} = 0.15331 \pm 0.00072 _{\rm stat} \pm 0.00040_{\rm
  syst}$ and $I_K^\mu|_{z/z} = 0.10101 \pm 0.00053 _{\rm stat} \pm
0.00039_{\rm syst}$.

%%%%%%%%%%%%%%%%%%%%%%%%%%%%%%%%%%%%%%%%%%%%%%%%%%%%%%%%%%%%%%%%%%
\section{Discussion}
\label{sec:discussion}
%%%%%%%%%%%%%%%%%%%%%%%%%%%%%%%%%%%%%%%%%%%%%%%%%%%%%%%%%%%%%%%%%%
While both the $z$ and dispersive parametrizations give rigorous 
bounds on theoretical uncertainties, the latter uses 
additional experimental input such as the low energy 
$K \pi$ phase shifts.  This allows for a one-parameter 
fit of the vector and scalar form factors, resulting in 
smaller uncertainties.
In the following, we discuss the impact of the value obtained
for $\ln C$, in the dispersive parametrization.

\subsection{Comparison with lattice results}

Here we compare our result for $\ln C$ against 
the lattice QCD calculations.  
This comparison does not depend on SM assumptions since no 
electroweak couplings are involved.

\begin{figure}[h!]
\begin{center}
\includegraphics[width=\linewidth]{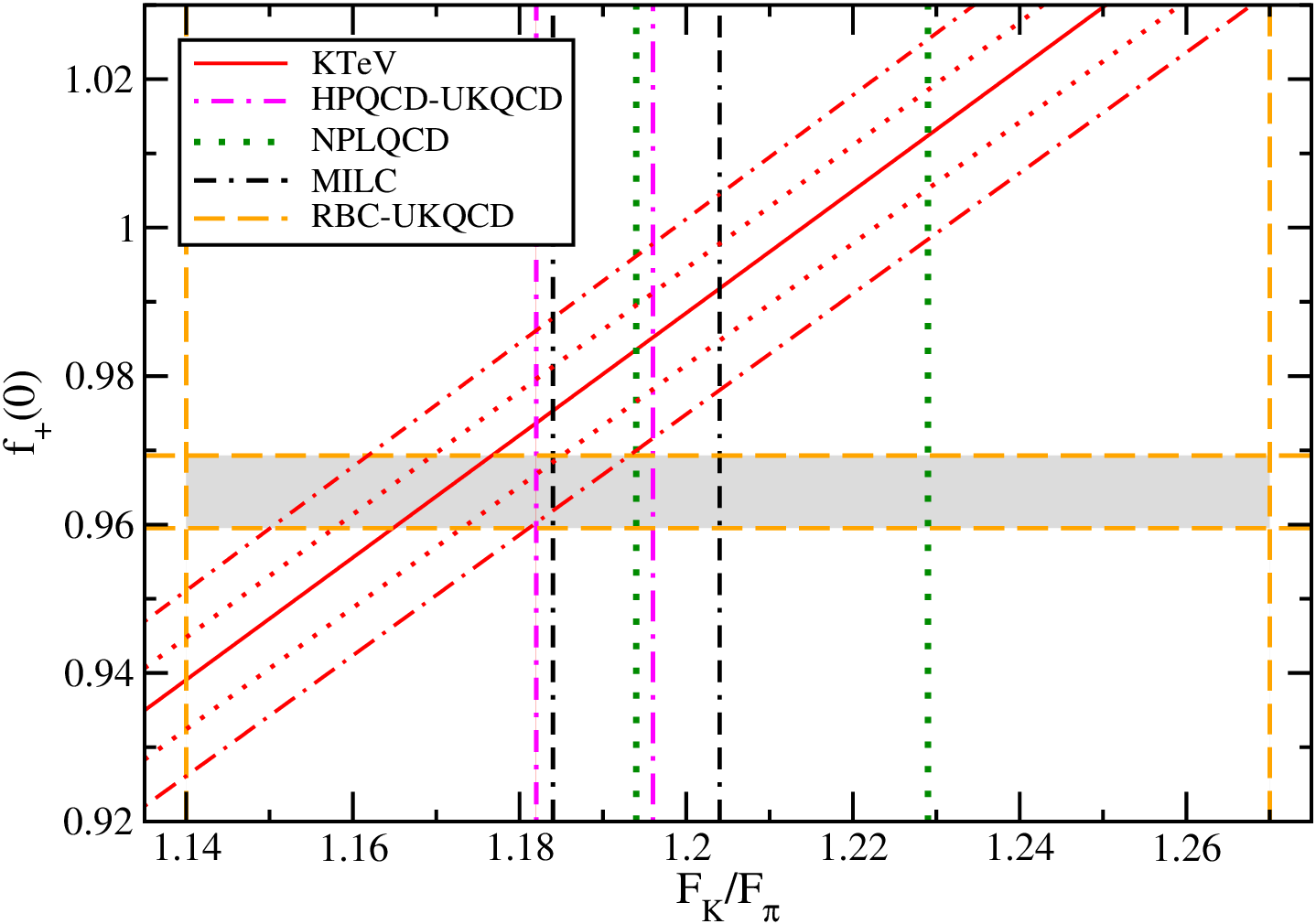}
\end{center}
\caption{{\it 
	The constraints on $f_+(0)$ and $F_K/F_\pi$ from various 
	lattice QCD calculations.  The KTeV result for 
	$F_K/(F_\pi \cdot f_+(0))$ is shown as the red solid line, 
	derived by using the Callan Treiman relation and the NLO 
	result for $\Delta_{CT}$.  Also shown is the error due to 
	$\Delta_{CT}$ (red dotted lines), and the resulting error 
	when added in quadrature to the total error 
    on $C$ as given in Eq. (\ref{eq:lnc}) (red dot-dashed line).}
	} % end caption
\label{fig:lattice}
\end{figure}

Figure~\ref{fig:lattice} shows  lattice QCD
results for $F_K/F_\pi$ and $f_+(0)$ in the 2+1 flavor case
\cite{milcfkfpi,hpqcdfkfpi,nplqcdfkfpi,rbcukqcdfkfpia,rbcukqcdfkfpib,ukqcdfplus}. 
A first classification of these data can be found in the recent
proceeding Ref. \cite{Lellouch09} awaiting for the FLAVIAnet
Lattice Averaging Group's one. We have only considered published results and showed them
as bands including systematic and statistical errors not giving the central values for clarity. 
Note that RBC/UKQCD has much bigger systematic errors compared to the other collaborations 
leading to the rather large band for $F_K/F_\pi$ ranging from 1.14 to 1.27.    

Also shown is the $f_+(0)$ vs $F_K/F_\pi$ dependence as derived from
Eq.~(\ref{C}) using our result for $\ln C$\footnote{Note that we have
  plotted $f_+(0)$ up to 1.02, though the Fubini Furlan bound
  \cite{flrs65} claims $f_+(0) \le 1$.  However, in QCD this bound can
  only be proven in the large $N_c$ limit, leaving the possibility for
  a slight deviation from it \cite{ste08}.}.  For $\Delta_{CT}$, we
have used the value~\cite{gale85}
\begin{equation}
 \Delta_{CT}=-0.0035 \pm 0.0080 \label{eq:dct},
\end{equation}
taken from a next-to-leading-order calculation in 
chiral perturbation theory in the isospin limit. 
The error is a conservative estimate of higher order corrections 
in the quark masses $m_{u}$, $m_d$ and $m_{s}$ \cite{leutpriv}.  
This value of $\Delta_{CT}$ 
is in agreement with other recent 
determinations~\cite{Bijnensdeltact,be08,JOP,kn08}.

%The lattice results for $F_K/F_\pi$ vary widely, leading to a range 
%between $1.14$ and $1.27
Combining all the lattice results, 
the grey band in the $f_+(0)$ vs $F_K/F_\pi$ plane   
shown  in Fig.~\ref{fig:lattice}, is obtained. 
Comparing this band to
the KTeV result, the ranges $F_K/F_\pi < 1.20$ and $f_+(0)>0.96$ 
are favoured by the KTeV data.  
%In particular, the KTeV result is 
%consistent with the last two lattice results by the 
%HPQCD-UKQCD \cite{hpqcdfkfpi} and RBC-UKQCD \cite{ukqcd07} 
%collaborations 
%that give ${F}_K/{F}_\pi=1.189(7)$ and $f_+(0)=0.9644(49)$,
%respectively.
%While these two lattice results have been favored by 
%the Flavianet Working Group \cite{flavia}, 
%we think that it is a bit premature to take these
%lattice results as definitive because the
%quoted errors do not include
%any estimate of the systematic uncertainties.  
%For a first attempt to include them, see Ref. \cite{rbcukqcdfkfpib}. 
%  Improving the experimental uncertainties on $\ln C$ together with
%  the lattice determination of $F_K/F_\pi$ and $f_+(0)$ could
%  certainly allow to improve our knowledge on these two important
%  quantities.}

\subsection{Test of the SM}
As mentioned in \S~\ref{sec:intro}, the small size of the
$\Delta_{CT}$ correction
allows for an accurate SM test using
the Callan Treiman relation, now rewritten as
\begin{equation}
\label{ctrel}
\frac{F_K}{F_\pi \cdot f_+(0)} = C-\Delta_{CT} .
\end{equation}
This test consists of comparing the value of ${F_K}/({F_\pi \cdot
  f_+(0)})$, deduced from the $K_{L\ell 3}$ dispersive form factor
parametrization fit, to the value of ${F_K}/({F_\pi \cdot 
f_+(0)})|_{SM}$, determined by assuming the SM electroweak (EW) 
couplings and using the experimental (photon inclusive) branching
fractions $\Gamma_{K^+_{\mu \nu}}/ \Gamma_{\pi^+_{\mu \nu}}$ and
$\Gamma_{K_{L e3}}$ measurements.

We define $r$ as
\begin{equation}
r = \left( C - \Delta_{CT} \right)\cdot \left( \frac{F_\pi \cdot f_+(0)}{F_K} \right) \Big{|}_{SM} ~ .
\end{equation}
Physics beyond the SM, such as modifications of EW couplings of
quarks due to new exchanges close to the TeV scale, could cause $r$ to
differ from unity.  
An example of modified EW couplings between right-handed quarks 
and the $W$ boson is discussed in Refs. \cite{bops06,bops07}.

We first calculate ${F_K}/({F_\pi \cdot f_+(0)})|_{SM}$.  
Assuming the SM couplings, one has
\begin{eqnarray}
\label{RPl2}
& & \frac{\Gamma[K\to l\nu_l(\gamma)]}{\Gamma[\pi\to
l\nu_l(\gamma)]}|_{SM} \nonumber \\
& & = {\rm cte}
\frac{|V_{us}|^2}{|V_{ud}|^2}\frac{F_K^2 m_K}{F_\pi^2
m_\pi}\frac{(1-x_K^2)^2}
{(1-x_\pi^2)^2}\frac{[1+\frac{\alpha}{\pi}F(x_K)]}{[1+\frac{\alpha}{\pi}
F(x_\pi)]} \nonumber \\
& & = \mathcal{M}^2 ~\left(\frac{F_K  V_{us}}{F_\pi V_{ud}} \right)^2~,
\end{eqnarray}
where $x_P\equiv m_l/M_P$.  The expression for cte, 
which depends on the hadronic structure and particle masses, 
can be found in Ref.~\cite{ms93}.  The function
$F(x)$ parametrizes the electromagnetic radiative corrections, 
and $\alpha$ is the fine structure constant.
% In Eq.~(\ref{RPl2}), we have isolated the quantity of interest, 
% namely the ratio of the kaon and pion decay constant times the 
% ratio of two CKM matrix elements.
The coefficient $\mathcal{M}$ thus defined is equal to 
0.2387(4) (see Ref.~\cite{bm06}).

The $K_{Le3}$ partial width is expressed as
\begin{equation}
\Gamma_{K_{Le3}}|_{SM} = \mathcal{N}^2 |f_+(0)V_{us}|^2,
\label{eq:ndef}
\end{equation}
where
\begin{equation}
\mathcal{N}^2 = G_F^{2}~\frac{m^5_K}{(192 \pi^3)}~S_{EW}~
 (1+ \delta^{e}_{K})~I^e_{K}. 
\end{equation}
% \begin{eqnarray}
% \Gamma_{K_{Le3}}|_{SM} &=& G_F^{2}~\frac{m^5_K}{(192 \pi^3)}~S_{EW}~
%  (1+ \delta^{e}_{K})~|f_+(0)V_{us}|^2~I^e_{K}   \nonumber \\
% &=& (\mathcal{N} |f_+(0)V_{us}|)^2~. 
% \label{eq:ndef}
% \end{eqnarray}
Here $G_F$ is the Fermi constant, $S_{EW}$ are short-distance
electroweak corrections, and $\delta_K^e$ denotes the 
electromagnetic (EM) radiative corrections.  
%
% Note that we have isolated on the right-hand side the quantity of 
% interest, $|f_+(0) V_{us}|$, and introduced a new constant 
% For convenience, we define $\mathcal{N}$ as
% \begin{equation}
% \mathcal{N}^2 = G_F^{2}~\frac{m^5_K}{(192 \pi^3)}~S_{EW}~
%  (1+ \delta^{e}_{K})~I^e_{K} 
% \end{equation}
%
From equations (\ref{RPl2}) and (\ref{eq:ndef}), it can be shown that
\begin{equation}
\label{bbb}
\frac{F_K}{F_\pi \cdot f_+(0)}\Big{|}_{SM} = 
\left( \frac{\Gamma_{K^+_{\mu \nu}}}{\Gamma_{\pi^+_{\mu \nu}} \cdot 
        \Gamma_{K_{Le3}}}\right)^{1/2}
\cdot |V_{ud}|  \cdot \frac{\mathcal{N}}{\mathcal{M}}.
\end{equation}

Using the world average result~\cite{bm06}
$\Gamma_{K^+ _{\mu \nu}}/ \Gamma_{\pi^+_{ \mu \nu}} = 1.3337(46)$, 
the KTeV measurement of 
$\Gamma_{K_{Le3}}=0.4067(11)$ \cite{KTeVe},
the values of 
$I_K^e$ from Table~\ref{tab:disp} and $\delta^e_K =0.0130(30)$ from
Ref.~\cite{andre}, 
and the value of $|V_{ud}|$ inferred from $0^+ \to 0^+$ 
superallowed nuclear transitions
\cite{ref:vud}\footnote{Note that a recent update \cite{ref:vudupdate} has appeared.},
\begin{equation}
|V_{ud}| =0.97418(26) ,
\label{Vud}
\end{equation}
Eq. (\ref{bbb}) gives the result
\begin{equation}
\frac{ F_K}{F_\pi \cdot f_+(0)} \Big{|}_{SM} = 1.2407 \pm 0.0044.
\label{smratio1}
\end{equation}

This result can be compared with the experimental
determination of $F_K/(F_\pi \cdot f_+(0))$ through $K_{L\ell 3}$
decay as given in Table 1. One obtains
\begin{equation}
r =1.0216 \pm 0.0124_{\rm exp} \pm 0.0039_{\rm theo} \pm 0.0067_{\Delta_{CT}}.
\end{equation}
The first two errors come from the experimental and the
theoretical uncertainties on $\ln C$ respectively,
and the last error comes from the estimated error on $\Delta_{CT}$
{\footnote{Note that this analysis of the KTeV data holds if there are
    no scalar, pseudoscalar and tensor couplings. Indeed in the
    presence of the two former what has been measured in table 1 is
    not $\ln C$ but $\ln C$ plus a small quantity, see Ref. \cite{flavia}.
    Consequently in that case a determination of $r$ is not possible
    without a precise knowledge of these couplings.}}. Adding
  the different errors in quadrature, we obtain $r = 1.0216 \pm
  0.0146$.

\begin{figure}[h!]
\begin{center}
\vspace{0.7cm}
\includegraphics[width=\linewidth]{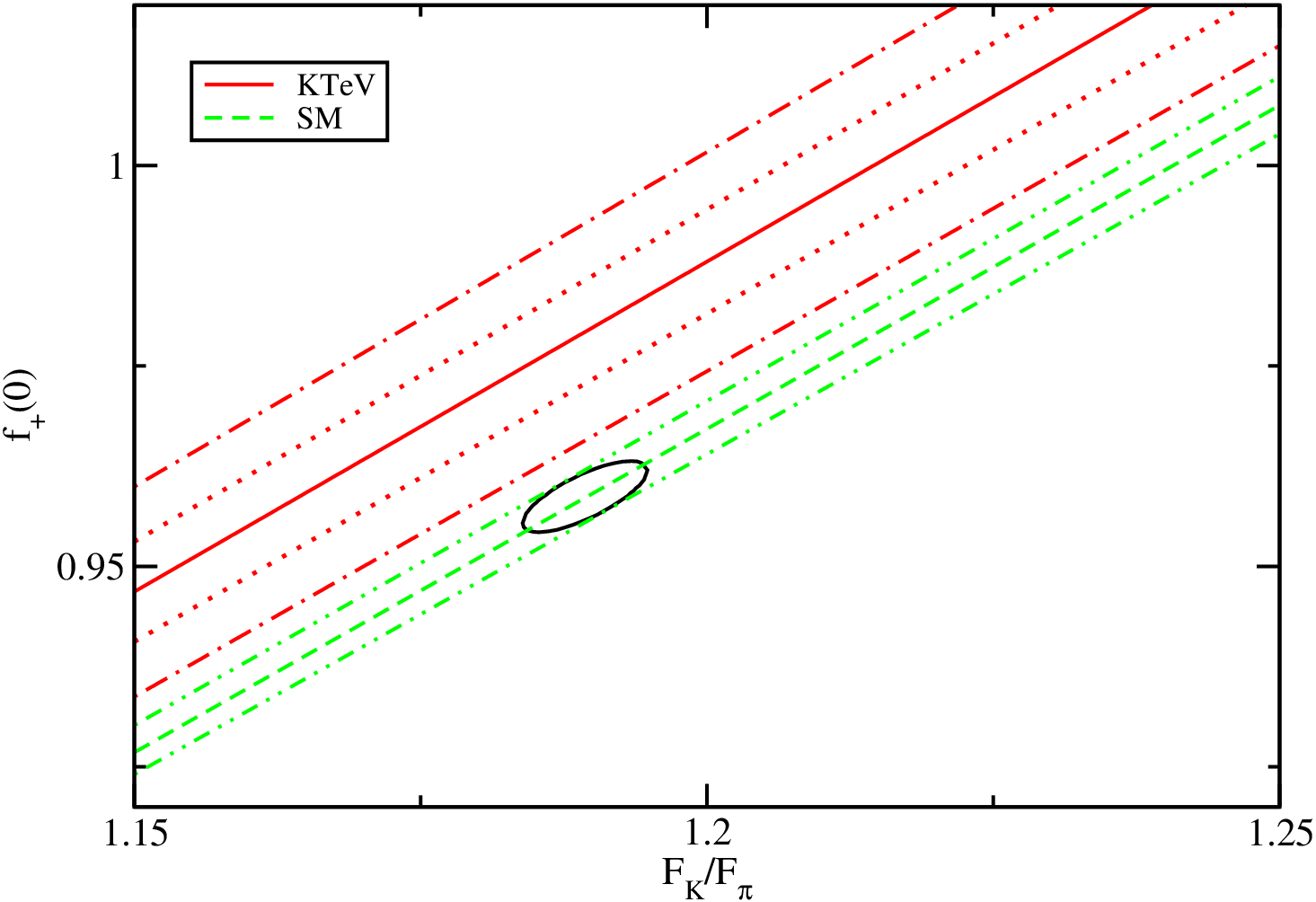}
\end{center}
\caption{{\it 
	The constraints on $f_+(0)$ and $F_K/F_\pi$. 
    The inserted $\chi^2 - \chi^2_{MIN} = 1 $ ellipse is 
	the SM result obtained from 
	Eqs. (\ref{RPl2}) and (\ref{eq:ndef}) and the CKM unitarity. 
%	and the smallness of 
%	$|V_{ub}|^2$ (Eq.~(\ref{hatFKFpi})). The dashed green
%	$|V_{ub}|^2$. 
	The dashed green 
	line and its error bar (green dotted line) corresponds 
	to $F_K/(F_\pi \cdot f_+(0))|_{SM}$, calculated using 
	the measured decay widths and assuming the SM couplings 
	of light quarks, Eq. (\ref{smratio1}). 
	Shown in red is the 
	$F_K/(F_\pi \cdot f_+(0))$ result, derived from the dispersive 
	parameterization fit to KTeV data and the NLO result for
	$\Delta_{CT}$.  Also shown is the error due to $\Delta_{CT}$ 
	(red dotted lines) added in quadrature to the total error 
    on $C$ as given in Eq. (\ref{eq:lnc}) (red dot-dashed line).}}  
%	The dashed green 
%	line and its error bar (green dotted line) corresponds 
%	to $F_K/(F_\pi \cdot f_+(0))|_{SM}$, calculated using 
%	the measured decay widths and assuming the SM couplings 
%	of light quarks, \textcolor{blue}{Eq. (\ref{smratio1})}. 
%	The inserted $\chi^2 - \chi^2_{MIN} = 1 $ 
%   ellipse is the {\textcolor{blue}{further
%	constraint}} given by CKM unitarity and the smallness of 
%	$|V_{ub}|^2$ (Eq.~(\ref{hatFKFpi})). Shown in red is the 
%	$F_K/(F_\pi \cdot f_+(0))$ result, derived from the dispersive 
%	parameterization fit to KTeV data and the NLO result for
%	$\Delta_{CT}$. Also shown is the error due to $\Delta_{CT}$ 
%	(red dotted lines), and the resultant error when added in 
%	quadrature to the experimental error (red dot-dashed line).}}
%	} % end caption
\label{treim}
\end{figure}

Analogous to Fig.~\ref{fig:lattice}, Fig.~\ref{treim} shows the two 
bands in the $f_+(0)$-vs-$F_K/F_\pi$ plane. The first band (red) 
shows the dispersive parametrization analysis of the KTeV $K_{L \ell 3}$
data. The second band (green) shows the SM prediction, 
Eq.~(\ref{smratio1}).
We observe a 1.5 $\sigma$  difference between the KTeV result and
the SM prediction.

Separate bounds on $F_K/F_\pi|_{SM}$ and $f_+(0)|_{SM}$ can be derived 
from CKM unitarity \cite{bops07, stern06}.  
Unitary implies that $|V_{ud}|^2+|V_{us}|^2+|V_{ub}|^2=1$, 
and measurements involving $b\to u$ transitions have shown that
$|V_{ub}|^2$ is negligibly small. 
Consequently, the SM mixing of light quarks is entirely
specified by the value of $|V_{ud}|$. %, Eq.~(\ref{Vud}).  

Substituting $|V_{us}|^2 = 1 - |V_{ud}|^2$ into equations 
(\ref{RPl2}) and (\ref{eq:ndef}), and solving for $F_K/F_\pi$ and $f_+(0)$, 
we obtain the contour in Fig. \ref{treim}. One has
% Therefore, Eq.~(\ref{RPl2}) together with the experimental
% value for $\Gamma_{K^+_{\mu\nu}}/\Gamma_{\pi^+_{\mu\nu}}$ 
% can be used to determine $F_K/F_\pi|_{SM}$. 
% The same is true for $f_+(0)|_{SM}$, which we can extract 
% assuming CKM unitarity from Eq.~(\ref{eq:ndef})
% and the value of $\Gamma_{K_{e3}}$.
%
%Following Ref.\cite{stern06} we will denote by a hat the values of
%the  QCD quantities $F_K$, $F_\pi$, $f_+(0)$ deduced in this way
%from (semi)leptonic decays assuming the SM  couplings of quarks. 
%
% Using the current experimental input from (semi)leptonic decays, we find
%we find with $|V_{ud}|$ from Eq. (\ref{Vud})
\begin{eqnarray}
F_K/F_\pi |_{SM} & = & 1.189 \pm 0.007 \nonumber \\
f_+(0) |_{SM} & = & 0.959 \pm 0.006. 
\label{hatFKFpi}
\end{eqnarray}
%
% A comparison of the SM prediction for $F_K/ F_\pi$ and $f_+(0)$ with
% the KTeV result is shown by the box
% in Fig.~\ref{treim}. 
%{\textcolor{blue}{Combining these equations with Eq. (\ref{smratio1}) 
%leads to the the black ellipse
%in Fig.~\ref{treim}.}} 
%{\textcolor{blue}{The corresponding bounds together with their correlation 
%are shown by the contour in Fig.~\ref{treim}.}}
%These unitarity-derived bounds are shown by the contour in Fig.~\ref{treim}. 
With the current experimental precision, 
the data show a marginal agreement with the SM
as concluded before.

\subsection{Ratio $G_F^\mu/G_F^e$}
\newcommand{\Rmue}{\mathcal{R}_{\mu/e}}

Taking the ratio of the $K_{Le 3}$ to $K_{L\mu 3}$ partial widths,
without assuming the equality of the $G_F^{\mu,e}$ decay constants, 
one obtains
\begin{equation}
\left(\frac{G^\mu_F}{G^e_F} \right)^2 =
\left[ \frac{\Gamma(K_L\to \pi^{\pm}\mu^{\mp}\nu)}
{\Gamma(K_L \to \pi^{\pm}e^{\mp}\nu)} 
\right]
/ 
\left(
\frac{1+\delta^{\mu}_K}{1+\delta^e_K} \cdot \frac{I^{\mu}_K}{I^e_K}
\right)~.
\end{equation}
Using the value of $I_K^{\mu}/I_K^e$
from Table 1, the EM correction estimates 
$(1+\delta^{\mu}_K)/(1+\delta^e_K)= 1.0058 \pm 0.0010$ 
from Ref. \cite{andre},\footnote{ 
To maintain consistency of this analysis with that of 
\cite{KTeVe}, 
the EM correction ratio has been taken from Ref.~\cite{andre}.   
Recently, new radiative correction estimates have appeared based 
on ChPT calculations \cite{cirneu}, which are in agreement 
with the old results but with a larger uncertainty.
} % end footnote
and the direct measurement of
$\Gamma(K_L\to \pi^{\pm}\mu^{\mp}\nu)/ 
 \Gamma(K_L \to \pi^{\pm}e^{\mp}\nu)=0.6640 \pm 0.0014 \pm 0.0022$ 
from  Ref. \cite{KTeVe}, one obtains
\begin{equation}
\Rmue \equiv \left(\frac{G^\mu_F}{G^e_F} \right)^2 = 0.9978 \pm 0.0049.
\label{mueuni}
\end{equation}
This result is in excellent agreement with the Standard Model 
expectation of unity, and it is very similar to the previous KTeV result,
$\left(G^\mu_F /G^e_F \right)^2 = 0.9969 \pm 0.0048$.
Although the $I_K^\mu$ and $I_K^e$ phase-space uncertainties 
in Table~\ref{tab:disp} are smaller than in the previous
KTeV analysis \cite{KTeVe},
the uncertainty on $\Rmue$ is almost identical in these two analysis. 
This is due to the fact that the uncertainties on 
the radiative corrections dominate the uncertainties on this ratio.
In both analyses, the uncertainty on the ratio of phase-space
integrals, $I_K^\mu/I_K^e$, is significantly smaller than the
quadrature-sum of the individual uncertainties because of
correlations in the vector form factor ($f_+$).
In the KTeV analysis, the theoretical uncertainty related to the 
scalar form factor ($f_0$) could not be evaluated because of the
parameterization used.
In this analysis using the dispersive parametrization, 
the $f_0$ uncertainty is more reliable, 
resulting in a more robust estimate of the
uncertainty on $\Rmue$.

\section{Summary}
\label{sec:summary}
A dispersive analysis of the semileptonic form factors for 
$K_{L e3}$ and $K_{L \mu 3}$ has been performed based on the 
published KTeV data.  

The measured value of $\ln C$, the scalar form factor at the
Callan-Treiman point, leads to a dependence of
$f_+(0)$ on $F_K/F_\pi$ within $1.5 \sigma$ 
of the Standard Model prediction.  
It favors an $F_K/F_\pi$ value on the lower
side of the lattice results, and an $f_+(0)$ value on the higher side.
% Using the dispersive parameterizations, 
New values of the decay phase
space integrals $I_K^\mu$ and $I_K^e$ are obtained, 
where the latter is consistent with the result obtained by 
$z$ parametrization~\cite{KTeVz}.
These new values can be used to determine the ratio of the decay constants 
of the two semileptonic modes, $G_F^\mu/G_F^e$, which is in excellent
agreement with the Standard Model prediction.
A detailed analysis of a Taylor-expansion vector form factor
fit to the data is used to study how the 
scalar and vector form factor correlations affect the result for $\ln C$.

\appendix

\section{Parameter Correlations}

In this Appendix, we will investigate correlations between parameters in the
different form factor parameterizations.
This study helps to understand 
several results discussed in the text, especially the 
difference between the parametrizations used for the vector form factor presented in 
Sec. \ref{sec:parametrisations}, and 
the robustness of the dispersive result presented in Sec. \ref{sec:analysis}. 

For this study, we will perform several fits of the $K_{L\mu3}$ data. We will always use the
dispersive parameterization for the scalar form factor. For the vector 
form factor, we will consider
a cubic expansion, i.e., the first three terms of Eq.~(\ref{Taylor}) will be
taken into account. 
Indeed, in the physical region of $K_{L\mu3}$ decay, a good representation 
of the dispersive parameterization may be obtained by Taylor expanding it 
with respect to $t$ and keeping only the first three terms. 
%Note that in the case of the vector form factor this
%third order coefficient turns out to be rather close to zero. 
We thus have four parameters ($\ln C$, 
$\lambda'_+$, $\lambda''_+$, and $\lambda'''_+$) 
which will enter the fitting procedure. 

\begin{figure}[h!]
\begin{center}
\vspace{0.7cm}
\includegraphics[width=\linewidth]{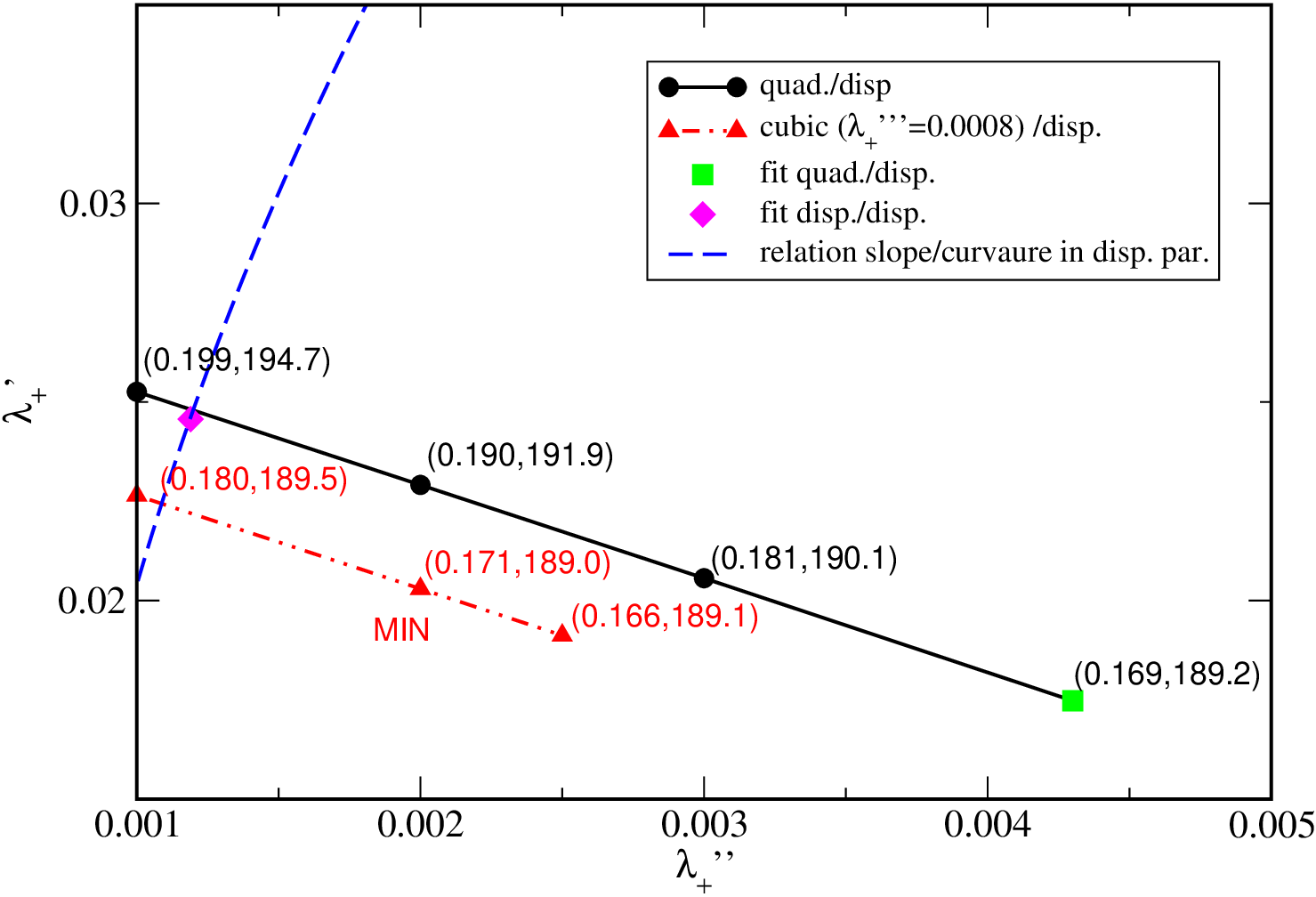}
\caption{\it Dependence of the fitted values of $\ln C$ and
  $\lambda_+'$ on the curvature, $\lambda_+''$. The value of 
  $\ln C$ and the $\chi^2$ of the corresponding fit are given in
  parenthesis above each point. To simulate the presence of higher
  order terms in the z-parameterization, for the red line a third order
  term has been included. The effect of this third order term is only
  to shift the curve downwards. In the legend, "quad./disp." for example indicates 
  that a "quadratic" parametrization is used for the vector form factor and 
  a "dispersive" one for the scalar.}
\label{fig:curvature}
\end{center}
\end{figure}

Let us first fix the third order coefficient $\lambda'''_+$ to zero, which
corresponds to a fit with a quadratic parametrization for the vector form factor.  We then perform a fit with two free parameters, 
namely $\ln C$ and $\lambda'_+$, while
fixing the curvature $\lambda_+''$ to three 
different values: $\lambda_+''= 0.001$, $0.002$ and $0.003$
($\lambda_+''=0.0$ would correspond to a linear parameterization for the vector form 
factor, which gives a very poor fit to the data). The results of these fits are
shown by the black points on Fig.~\ref{fig:curvature}. 
%%%%%%%%%%%%%%%%%%%%%%%%%%%%%%%%%%%%%%%%%%%%%%%%%%%%%%%
%%%%%%%%%%%%%%%%%%%%%%%%%%%%%%%%%%%%%%%%%%%%%%%%%
The value of $\lambda'_+$ can be read on the $y$ axis 
and $\ln C$ is given by the first number in parentheses above 
each black point; 
the second number in parentheses gives the $\chi^2$ of the fit.
The linear relationship between  $\lambda_+'$ and $\lambda_+''$ seen on the figure 
shows the strong correlation between these two quantities.
For example, a fit
to $K_{e3}$ data using a quadratic parameterization for the vector
form factor leads to a correlation,
$\rho(\lambda_+',\lambda_+'')=-0.96$~\cite{KTeVe}. A similar linear
dependence is also obtained between $\ln C$ and $\lambda_+'$ or
$\lambda_+''$. 
Furthermore, when moving toward larger $\lambda_+''$, $\ln C$ 
decreases while the $\chi^2$ decreases only slightly. The green square at the end
of the solid black line represents the minimum $\chi^2$; it corresponds to the result of the 
quadratic/dispersive fit given in Section~\ref{sec:parametrisations}. 
(We have stopped the line at that value of the curvature for convenience.)

Next, we consider the impact of a third-order term in the Taylor expansion. 
Indeed, a dispersive or z-parameterization of the vector form factor gives 
a non-zero value for the third order coefficient $\lambda_+'''$ 
when Taylor expanded. 
We repeat 
the same procedure as above (i.e., floating $\ln C$ and $\lambda'_+$, 
and varying $\lambda_+''$), but the third-order coefficient is now fixed to $\lambda_+'''=0.0008$, 
close to what is obtained from the results of the z-parameterization
(Tables II and III). 
The nonzero $\lambda_+'''$ causes 
an offset in the linear dependence of $\lambda_+'$ with $\lambda_+''$, 
but leads to the same slope and 
a similar variation of $\ln C$ with the slope and the curvature (see 
the red
dot-dot-dashed line on Fig.~\ref{fig:curvature}) 
\footnote{Note that for illustration purpose we 
only show a portion of this line as well as of the black solid line 
corresponding to the fits we have made. They of course go on in both 
directions.}. The minimum $\chi^2$ fit 
(green square for $\lambda_+'''=0$) is now moved toward a 
smaller $\lambda_+''$ and a larger $\lambda_+'$ 
(red triangle denoted by 'MIN'. This point corresponds to 
the z-parameterization result in Table II.) 
The parameter shifts are required in order to compensate for 
the additional cubic term. 
Note that the value of $\ln C$ is roughly the same at this minimum as
the one obtained previously for $\lambda_+'''=0$.

In the light of this study, let us now consider the other parameterizations 
used in the literature \cite{KTeVe}, \cite{KTeVz}, \cite{KLOE}, \cite{NA48mu}. 
One can distinguish two classes. 
One class (Class I), of which the dispersive and pole parameterizations are examples,
impose physically
motivated  relations between the slope and the
curvature (and possibly the higher order terms in the Taylor
expansion, the third order term being the most relevant one in the
physical region). Indeed, as already emphasized, in the dispersive
parameterization the curvature and all higher order terms
of the vector form factor are constrained not only on first principles, 
such as
analyticity and unitarity, but also by including the information on the
low energy $K \pi$ $P$-waves ($K^*$ resonance).  The relation between
slope and curvature is illustrated by the blue dashed
curve in Fig.~\ref{fig:curvature}. This curve 
crosses the black solid line and the red dot-dot-dashed line
for large values of $\lambda_+'$, small values of $\lambda_+''$, and
consequently large values for $\ln C$, nicely illustrating the result
of the dispersive/dispersive fit (magenta diamond in the figure).  
In contrast, the
second class (Class II), 
of which the Taylor series and z-parameterization are examples,
is based on mathematically rigorous
expansions, in which the slope and curvature are free parameters.  
Clearly, the
existence of a relation between slope and curvature strongly
constrains the fit in Class I,  while the fit has more freedom when using Class II parameterizations.
One thus expects smaller fitting 
uncertainties in the results from the Class I and larger ones from Class
II, so that if the theoretical errors are well controlled, 
the overall uncertainty will be smaller for class I.

% ====================================================
\vspace{0.2cm}
\hspace{-0.5cm}{\bf \large{ Acknowledgments}}

One of us (V. B.) would like to thank S. Descotes-Genon and 
B. Moussallam for interesting
discussions. This work has been partially supported by the EU contract
MRTN-CT-2006-035482 (``Flavianet''), the EU Integrated Infrastructure
Initiative Hadron Physics (RH3-CT-2004-506078), the Swiss National 
Science Foundation and the IN2P3 projet
th\'eorie ``Signature exp\'erimentale des couplages \'electrofaibles non-standards  des quarks''.

\end{document}